\documentclass[12pt]{iopart}

\usepackage[sorting=none]{biblatex}
\usepackage[pdftex]{graphicx}
\usepackage{bm}
\usepackage{nicefrac}
\usepackage{amssymb}

\begin{document}
\title{Biskyrmion-based artificial neuron}

\author{Ismael Ribeiro de Assis}
\address{Institut f\"ur Physik, Martin-Luther-Universit\"at Halle-Wittenberg, D-06099 Halle (Saale), Germany}

\author{Ingrid Mertig}
\address{Institut f\"ur Physik, Martin-Luther-Universit\"at Halle-Wittenberg, D-06099 Halle (Saale), Germany}

\author{B{\"o}rge G{\"o}bel}
\address{Institut f\"ur Physik, Martin-Luther-Universit\"at Halle-Wittenberg, D-06099 Halle (Saale), Germany}
\ead{boerge.goebel@physik.uni-halle.de. Corresponding author}
\vspace{10pt}
\begin{indented}
\item[]November 2022
\end{indented}

\begin{abstract}
Magnetic skyrmions are nanoscale magnetic whirls that are highly stable and can be moved by currents which has led to the prediction of a skyrmion-based artificial neuron device with leak-integrate-fire functionality. However, so far, these devices lack a refractory process, estimated to be crucial for neuronal dynamics. 
Here we demonstrate that a biskyrmion-based artificial neuron overcomes this insufficiency. When driven by spin-orbit torques, a single biskyrmion splits into two subskyrmions that move towards a designated location and can be detected electrically, resembling the excitation process of a neuron that fires ultimately. The attractive interaction of the two skyrmions leads to a unique trajectory: Once they reach the detector area, they automatically return to the center to reform the biskyrmion but on a different path. During this reset period, the neuron cannot fire again. Our suggested device resembles a biological neuron with the leak, integrate, fire and refractory characteristics increasing the bio-fidelity of current skyrmion-based devices.
\end{abstract}

\newpage
\section{Introduction}
The field of magnetism has been crucial for developing technological devices over the last decades. As an example, data can be stored by encoding bits via magnetic domains. An established strategy to improve such storage concepts is to decrease the size of these information carriers.~\cite{bobeck1969magnetic,michaelis1975magnetic,parkin2004shiftable}. Here, magnetic skyrmions~\cite{bogdanov1989thermodynamically,muhlbauer2009skyrmion,yu2010real} are promising candidates~\cite{sampaio2013nucleation, fert2013skyrmions}. These magnetic whirls possess topological properties allowing for stability even on the nanometer scale~\cite{heinze2011spontaneous,nagaosa2013topological} and they can even be moved by currents~\cite{jonietz2010spin,jiang2017direct,litzius2017skyrmion}.

Besides their potential for conventional spintronic devices, skyrmions are attractive for neuromorphic computing~\cite{grollier2016spintronic,li2017magnetic,huang2017magnetic,song2020skyrmion,li2021magnetic, azam2018resonate, Bindal_2022}. This promising approach is oriented at our brain's operational mode. We can solve complex tasks like face recognition at a fraction of computers' power consumption when they use conventional algorithms~\cite{lawrence1997face, markovic2020physics}. This is possible because neurons are connected via synapses and exchange electrical signals such that various input stimuli lead to a specific response. After training, most often, the output is sensible even though the underlying mechanism may be difficult to understand. Artificial neural networks mimic this behavior and can perform complex tasks  with ease compared to conventional computer algorithms.

\begin{figure*}[t!]
    \centering
    \includegraphics[width=\textwidth]{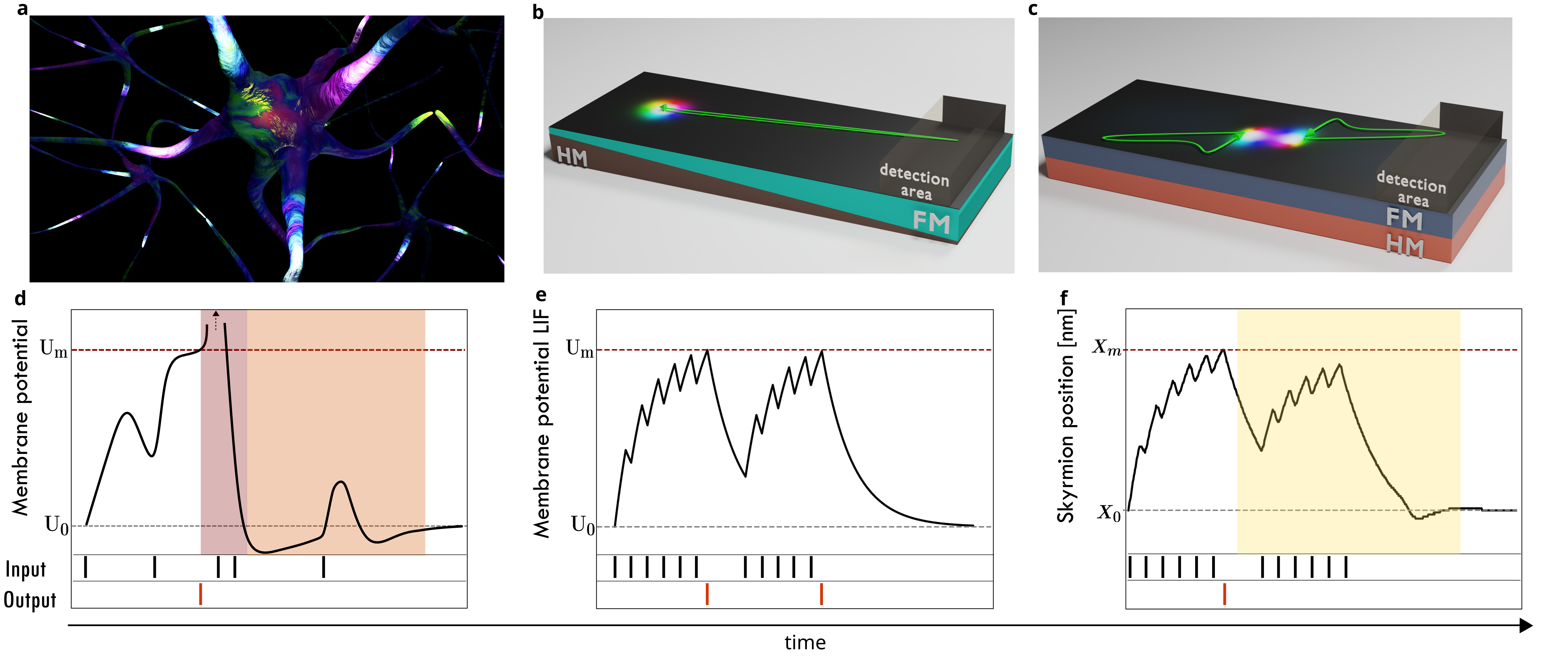}
    \caption{Overview of biological and artificial neurons. (a) Network of biological neurons. Input and output signals (highlighted regions) travel along the neuron body. (b) Schematic figure of the skyrmion-based artificial neuron similar to the one proposed in reference~\cite{li2017magnetic}. A skyrmion (colored object) in a ferromagnetic layer (FM) is driven on the green trajectory towards the detection area by spin-orbit torques caused by the heavy metal layer (HM). (c) Biskyrmion-based artificial neuron proposed in this work. The biskyrmion splits under current pulses (two green trajectories) and the subskyrmions can be detected. The skyrmions come back afterwards without the need of a wedge geometry. (d) Schematic figure of the membrane potential of a biological neuron with indicated input and output current pulses. The neuron fires once the potential reaches $U_m$ and enters an absolute refractory period (red area). Upon resetting towards $U_0$ the neuron overshoots and enters a relative refractory period (orange area). (e) Membrane potential of an artificial neuron that only has the LIF characteristics but no refractory property, like the skyrmion-based neuron in b. (f) Biskyrmion-based neuron characteristics corresponding to c. The device has the LIF properties plus a refractory period (yellow) during which it will not fire again no matter how many input pulses are applied.}
    \label{fig:fig1_neuron_devices}
\end{figure*}

Neural networks consist of neurons -- cells that non-linearly translate sequences of input currents into a response -- and synapses that weigh the current pulses. Like a biological neuron [figure~\ref{fig:fig1_neuron_devices}(a)], an artificial neuron should fulfill the following characteristics: integrate, fire, leak and refractoriness~\cite{gerstner2014neuronal}.
The essential dynamics of a biological neuron~\cite{zhu2020comprehensive} are depicted schematically in figure~\ref{fig:fig1_neuron_devices}(d):
A postsynaptic neuron receives current pulses via synapses (input: black lines). The membrane potential increases (i.\,e. the integrate characteristics) but drops towards its residual value $U_0$ without input signals (leak). If the membrane reaches a threshold value $U_m$ with the arrival of multiple input spikes, the neuron fires [output: red line in figure~\ref{fig:fig1_neuron_devices}(d)]. After firing, the membrane potential returns to $U_0$. During this process, the neuron is unable or inhibited from firing again (refractoriness).

An artificial neuron based on a magnetic skyrmion in a wedge [figure~\ref{fig:fig1_neuron_devices}(b)] has been proposed in reference~\cite{li2017magnetic}. Current pulses move the skyrmion (integrate) towards a detector, where the neuron fires, and the wedge geometry resets the skyrmion in the absence of currents (leak). However, like most other artificial neurons~\cite{zhu2020comprehensive}, this idea lacks the fundamental concept of a refractory period meaning that the neuron would continue to fire when current pulses are received in fast sequence [figure~\ref{fig:fig1_neuron_devices}(e)], like in a LIF (leak-integrate-fire) model~\cite{gerstner2014neuronal}; see Appendix for details of this simulation. A schematic animation of a LIF-based neuron is provided in the supplementary information (SI).

Here, we present an artificial neuron based on a magnetic biskyrmion [figure~\ref{fig:fig1_neuron_devices}(c)]. This object consist of two subskyrmions that are stabilized by dipole-dipole interactions~\cite{gobel2019forming,capic2019stabilty,capic2019biskyrmion}. The biskyrmion has been observed in centrosymmetric materials~\cite{yu2014biskyrmion,wang2016centrosymmetric,peng2017real,zuo2018direct,gobel2021beyond}. As we show, the two skyrmions separate when influenced by spin currents which allows for unique trajectories automatically leading to refractory periods during which the neuron cannot fire again [figure~\ref{fig:fig1_neuron_devices}(f)].

We present micromagnetic simulations and show that the skyrmions move along opposite directions towards a detector when driven by spin currents. Once the detector is reached by one skyrmion, the neuron fires. Due to the topological properties of the skyrmions, they move at an angle towards the edge~\cite{zang2011dynamics,jiang2017direct,litzius2017skyrmion}. Once the edge stops the forward motion, the skyrmions alter their direction of velocity driven by the weak attractive interaction between the two skyrmions [figure~\ref{fig:fig1_neuron_devices}(c)]. The neuron cannot fire again until the skyrmions reestablish the biskyrmion, ultimately resetting the neuron. We explain this unusual trajectory based on the Thiele equation~\cite{thiele1973steady}. 
This artificial neuron exceeds the capabilities of the typical LIF neuron by inherently incorporating a refractory process. For this reason, we believe that the biskyrmion-based neuron will be highly relevant for developing skyrmion-based neuromorphic technologies.

\section{Simulated system \& Methods}

We have conducted micromagnetic simulations and started with a single biskyrmion in a rectangular magnetic film [figure~\ref{fig:fig1_neuron_devices}(c)]. The rectangular magnetic layer is interfaced with a heavy metal, as shown in figure~\ref{fig:fig1_neuron_devices}(c), such that we can manipulate the biskyrmion via spin-orbit torque (SOT): When an electric current (pulse) $j$ is applied along $x$, the spin Hall effect~\cite{kato2004observation} generates a spin current along $z$ with spins $\bm{s}$ oriented along $y$. These spins interact with the magnetic moments of the biskyrmion and lead to a motion of the two subskyrmions. The used parameters are given below. However, the following observations are not restricted to one particular centrosymmetric material but remain valid for other sets of parameters as long as a biskyrmion can be stabilized.

We track the position of the skyrmion moving towards the detector [the skyrmion moving to the right in figure~\ref{fig:fig1_neuron_devices}(c)]. Once the skyrmion core reaches the detector, e.\,g. a magnetic tunneling junction, the device fires because a perpendicular current can flow. The skyrmion-skyrmion interaction then resets the device, with a  complete reset corresponding to the reformation of the biskyrmion. 

We have used the GPU accelerated software package Mumax3~\cite{vansteenkiste2011mumax,vansteenkiste2014design} to solve the LLG equation with the SOT term. The equation for the discretized magnetization reads
\begin{equation}
\partial_t \mathbf{m}_i = -\gamma_e \mathbf{m}_i \times \mathbf{B}^{i}_\mathrm{eff} + \alpha  \mathbf{m}_i \times  \partial_t \mathbf{m}_i + \gamma_e \epsilon \beta[( \mathbf{m}_i \times \mathbf{s}) \times \mathbf{m}_i] \label{eq:LLG_mumax3}
\end{equation}
where $\mathbf{B}^{i}_\mathrm{eff} = \delta F/M_s \delta \mathbf{m}_i$ is the effective field derived from the system's total free energy density $F$, given as the sum of exchange interaction, magnetocrystaline anisotropy, Zeeman energy and the demagnetization field (dipole-dipole interaction) -- the main interaction responsible for the stabilization of the biskyrmion. The constants in equation~\ref{eq:LLG_mumax3} are: the gyromagnetic ration $\gamma_e = 1.760 \times 10^{11} $ T$^{-1}$s$^{-1}$ and $\epsilon \beta= \frac{\hbar \Theta_{SH}}{2 e d_z M_s}$; where $d_z$ is the thickness of the magnetic layer, $e$ the electron's charge, $\hbar$ Planck's constant, $M_s$ the saturation magnetization and $\Theta_\mathrm{SH} j$ the spin current with spin orientation $\mathbf{s}$ generated by the spin-Hall angle $\Theta_\mathrm{SH}$.

The FM layer in figure~\ref{fig:fig1_neuron_devices}(c) is discretized in cells of size $1$ nm$\times$ $1$ nm $\times$ $1$ nm. The ferromagnetic orientation points along $-z$. The simulated parameter are: thickness $d_z = 3$ nm, Gilbert damping parameter $\alpha = 0.3$, saturation magnetization $M_s = 1.4$ MA/m, exchange constant $A = 15$ pJ/m, uniaxial anisotropy $K_z = 1.2$ MJ/m$^3$ and the external field $ B_z = -40$ mT, as in reference~\cite{gobel2019forming}.
To stabilize the biskyrmion, we have simulated the method proposed in reference~\cite{gobel2019forming}. Two individual Bloch-skyrmions with opposite helicities are written $64\,\mathrm{nm}$ apart in the ferromagnetic layer. Their attractive interaction leads to the formation of a biskyrmion.

\section{Results \& Discussion}
The refractory property of the biskyrmion-based artificial neuron that we predict in this paper is based on the unique trajectory of the two subskyrmions that form the magnetic biskyrmion: Their path towards the detector is different from the path back to the initial biskyrmion state. To characterize and understand this fascinating dynamics we first discuss it under constant current and afterward under the influence of current pulses, like in an artificial neuron device.

\subsection{Skyrmion pair driven by constant spin-orbit torque in micromagnetic simulations.}
A biskyrmion consists of two circular subskyrmions (figure~\ref{fig:fig2_biskyrmion}) both of which are magnetized along the out-of-plane direction in their center, opposite to the magnetization direction of the surrounding. In between, the magnetization $\bm{m}(\bm{r})$ has an in-plane component. Since the two skyrmions overlap partially (middle of figure~\ref{fig:fig2_biskyrmion}), their helicity, characterizing the in-plane profile, must differ by $\pi$. That means two distinct types of skyrmions must form a biskyrmion, which is why a biskyrmion is disfavored by the Dzyaloshinskii-Moriya interaction and is instead stabilized by the dipole-dipole interactions that favor Bloch skyrmions of helicity $+\nicefrac{\pi}{2}$ and $-\nicefrac{\pi}{2}$, likewise~\cite{gobel2019forming}.

The two subskyrmions of the biskyrmion behave differently when we drive them by spin-orbit torques, due to their opposite helicity and corresponding in-plane magnetization profiles. For positive currents that are larger than a critical current, the two skyrmions move away from each other; see orange trajectory in figure~\ref{fig:fig3_motionSOT}(a). However, they do not move perfectly (anti-)parallel with respect to the current direction but move at an angle partially towards the edge of the sample. This effect is called skyrmion Hall effect~\cite{zang2011dynamics,jiang2017direct,litzius2017skyrmion} and will be further clarified in the next section based on the Thiele equation~\cite{thiele1973steady}. Once the two skyrmions reach their respective edge after $9\,\mathrm{ns}$, the motion almost stops. However, both skyrmions then begin to creep along the edge towards each other along the $\pm x$ direction. The motion along the $x$ direction has reversed even though the current remains unchanged. This motion is much slower compared to the initial separation process and after a total $180\,\mathrm{ns}$ the two skyrmions have the same $x=0$ component but are still positioned at the opposite edges with respect to the $y$ coordinate. This configuration is a steady state under the applied constant current. However, once the current is turned off, the two skyrmions attract each other again and merge to reestablish the initial biskyrmion. Note that this will occur automatically in the neuronal operation mode where the input current is received pulsed.

Before we continue and explain this unique trajectory in detail we want to comment on three alternative scenarios that might occur in practice. First, even if the current is turned off at any other point of the trajectory, the two skyrmions still attract each other and form the initial biskyrmion. Second, if the current has the wrong sign (or if the two subskyrmions are reversed) the biskyrmion first rotates by $\pi$, effectively exchanging the two subskyrmions. Third, if the driving current is too small, the biskyrmion will only rotate and the two individual skyrmions do not form. Therefore, as long as the driving current is large enough, these scenarios are unproblematic and it is sufficient to focus on explaining the trajectory described above.

\subsection{Explanation of the non-linear motion using the Thiele equation.} 
\begin{figure}[tb]
    \centering
    \includegraphics[width=0.5\columnwidth]{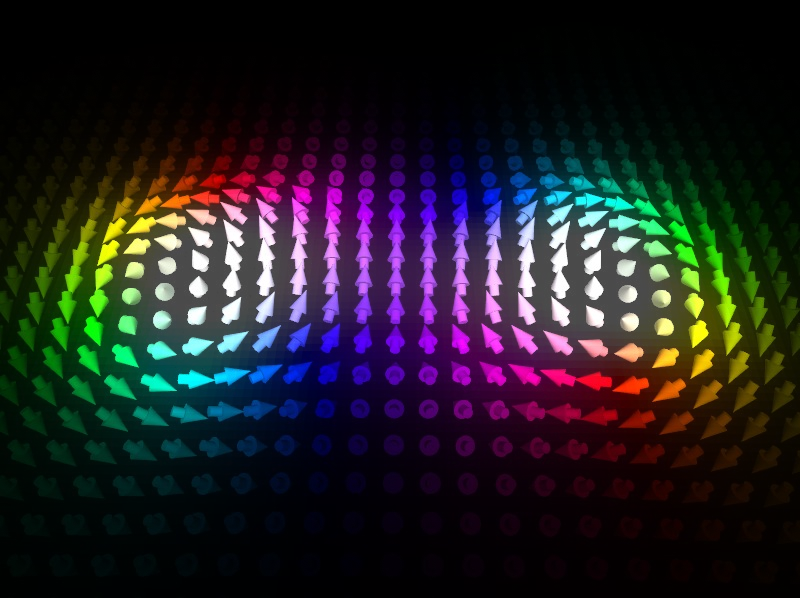}
    \caption{Magnetic biskyrmion. Each arrow represents the magnetization in the respective cell in our micromagnetic simulation. Black regions consist of black arrows pointing along $-z$. White arrows indicate an orientation along $+z$ and the color encodes the polar angle in the $xy$ plane.}
    \label{fig:fig2_biskyrmion}
\end{figure}

To understand the biskyrmion motion via SOT, we use the generalized Thiele equation~\cite{thiele1973steady}. It is an effective description of the motion of non-collinear textures with the velocity $\bm{v}$. The essential assumption is that the (bi-)skyrmion spin texture does not change in profile while moving, so that it can be condensed into a single point, and that the total force in the system vanishes. Since we know from the micromagnetic simulations that the biskyrmion splits up into two subskyrmions, we continue to analyze the trajectory of the skyrmions. The right skyrmion [figure~\ref{fig:fig2_biskyrmion} and figure~\ref{fig:fig1_neuron_devices}(c)] is discussed if not stated otherwise because we will consider this skyrmion for the detection later on.

The Thiele equation consists of five force terms and can be written as~\cite{gobel2019overcoming}
\begin{equation}
    b \bm{G} \times \bm{v} - b \underline{D} \alpha \bm{v} - B j \underline{I} \bm{s} - \nabla U_{\mathrm{int}}(r_{1,2}) - \nabla U_{\mathrm{edge}}(y) = 0.
    \label{eq:Thiele}
\end{equation}
The constants $b$ and $B$ are determined from the sample parameters (see Methods section): $b = M_s d_z / \gamma_e$ and $ B = \hbar \Theta_\mathrm{SH} / 2 e $ where $M_s$ is the saturation magnetization, $d_z$ is the thickness of the ferromagnetic sample, $\gamma_e$ is the gyromagnetic ratio and $\Theta_\mathrm{SH}$ is the spin Hall angle.

The topological charge of the (bi-)skyrmion 
$N_\mathrm{Sk} = \frac{1}{4 \pi} \int \bm{m} \cdot (\partial_x \bm{m} \times \partial_y \bm{m} ) \, \mathrm{d}^2 r$
gives rise to the first term. This so-called gyroscopic force is characterized by the gyroscopic vector $\bm{G} = 4 \pi N_\mathrm{Sk}$. The space-dependent magnetic profile $\bm{m}(\bm{r})$ has been condensed into a single vector by integrating over the whole extent of the skyrmion. Each skyrmion has a topological charge of $N_\mathrm{Sk}=+1$. The second term is the dissipative force, quantified by the Gilbert damping $\alpha$. The dissipative tensor 
$D_{ij} = \int ( \partial_{x_i} \bm{m} \cdot \partial_{x_j} \bm{m} ) \, \mathrm{d}^2 r$
only has non-zero $D_{xx}=D_{yy}\equiv D_0$ elements, irrespective of the type of skyrmion, as long as there is no deformation. The third term accounts for the spin-orbit torque. The injected spins $\bm{s}$ interact with the skyrmion's magnetic moments. The torque tensor 
$I_{ij} = \int ( \partial_{x_i} \bm{m} \times \bm{m} )_j \,\mathrm{d}^2 r$
strongly depends on the skyrmion's in-plane magnetization profile; more precisely on its helicity. For the two Bloch skyrmions with opposite helicity $\gamma=\pm\nicefrac{\pi}{2}$ only the $I_{xx}=I_{yy}\equiv \lambda I_0$ components are non-zero, as long as there is no deformation. Note that $\lambda=\pm 1$ has been introduced to distinguish the two skyrmions with positve and negative helicity, respectively.

The other two terms are the interaction between the two skyrmions, quantified by $U_\mathrm{int}(r_{12})$, and the interaction of a skyrmion with the edge $U_\mathrm{edge}(y)$. If we neglect them, for now, we are able to understand why the two skyrmions move away from each other once the current is turned on. Under the above explained symmetry considerations, the Thiele equation becomes
\begin{equation}
0=-4\pi b \left( \matrix{-v_y\cr\\v_x\cr} \right)-bD_0\alpha\left( \matrix{v_x\cr\\v_y\cr} \right)- \lambda BjI_0\left( \matrix{0\cr\\1\cr} \right) \label{eq:Thiele_vector}
\end{equation}
Both skyrmions move at the skyrmion Hall angle $\tan\theta_\mathrm{sk}=\nicefrac{v_y}{v_x}=\nicefrac{D_0\alpha}{4\pi}$ with respect to the current direction $x$ and they move along opposite directions so that the biskyrmion splits like we have seen in the micromagnetic simulations.

\begin{figure}[tb]
    \centering
    \includegraphics[width=0.8\columnwidth]{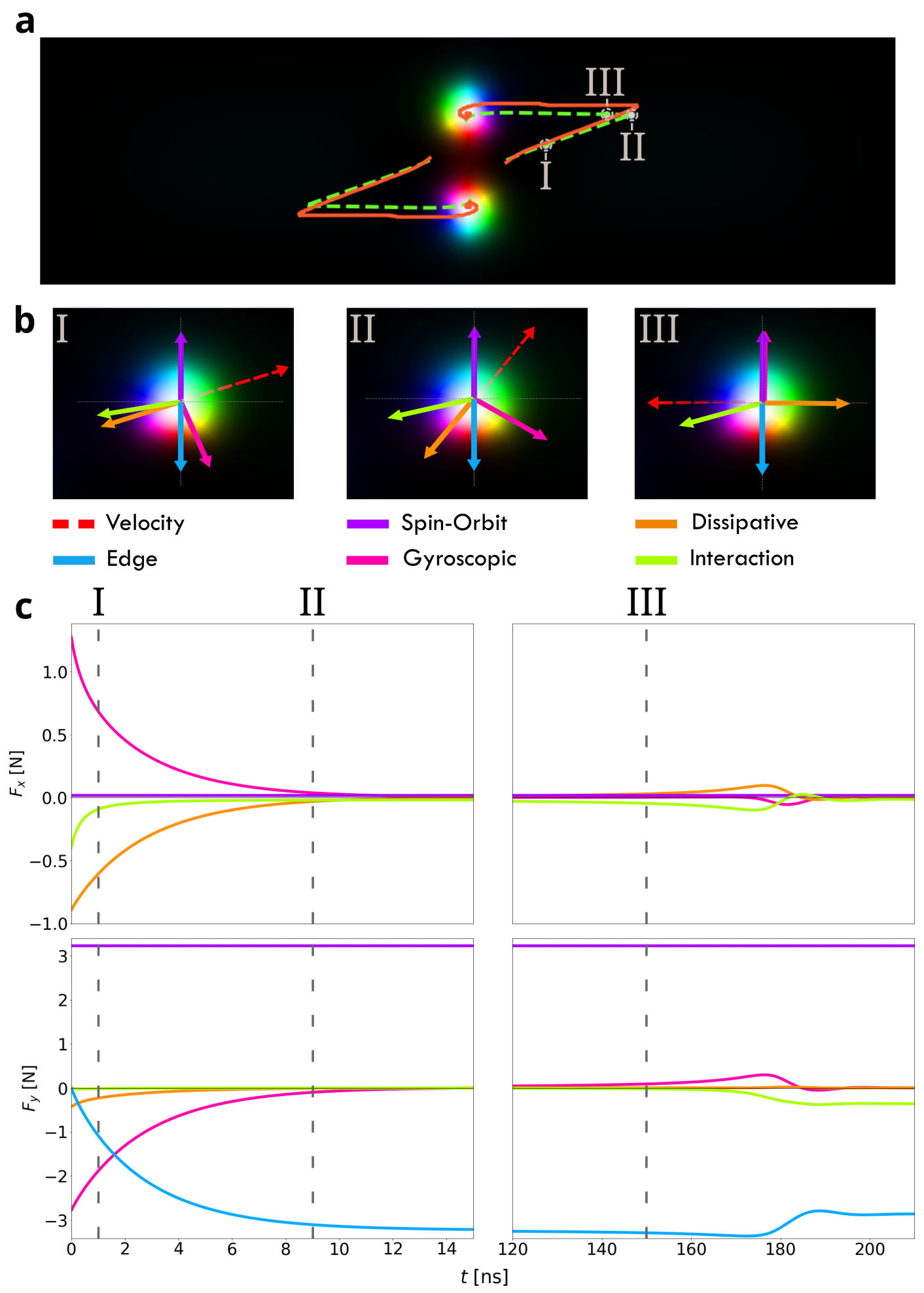}
    \caption{Skyrmion motion under spin-orbit torque. (a) Trajectories from the micromagnetic simulations (orange) and the Thiele equation (dashed green). An animated version is available in the SI. (b) Snapshots of the right skyrmion as indicated by I, II, III in a. The arrows represent the orientation of the velocity (dashed) and the five force terms entering the Thiele equation (solid), as indicated. Due to considerable differences in length, the arrows have been normalized to a fixed length for a clearer representation. (c) The magnitude of the forces corresponding to the arrows in b. The time range between $14\,\mathrm{ns}$ and $120\,\mathrm{ns}$ has been omitted because the forces are almost zero in this case.}
    \label{fig:fig3_motionSOT}
\end{figure}

The two skyrmions move at that angle away from each other until they approach their respective horizontal edge. This leads to a force $\nabla U_\mathrm{edge}$ along $\lambda y$. Since this force does not have an $x$ component, the first component of equation~\ref{eq:Thiele_vector} reveals that the skyrmions can still only move at the skyrmion Hall angle. This also means that once the force from the potential compensates the forces from the spin-orbit torque at $y=y_c$, the skyrmion cannot move anymore at all $v_x=v_y=0$. Note that if the right skyrmion was somehow displaced beyond $y>y_c$, it would move along the opposite direction but along the skyrmion Hall angle until it reaches $y=y_c$. This is a special feature of Bloch skyrmions compared to N\'eel skyrmions, which have a different $\underline{I}$ tensor symmetry. While N\'eel skyrmions can creep along the edges of a confined geometry~\cite{iwasaki2013current}, Bloch skyrmions always get stuck. 

The only remaining force we have not discussed yet, is the skyrmion-skyrmion interaction which is attractive for all points of the observed skyrmion trajectory. Even though it is weak compared to all other interactions, it has an $x$ component which allows the two skyrmions to leave the straight course dictated by the skyrmion Hall angle.

We qualitatively reproduced the unique trajectory [figure~\ref{fig:fig3_motionSOT}(a)] by solving the Thiele equation after calculating and fitting $D_0$, $I_0$, $U_\mathrm{int}(\bm{r})$ and $U_\mathrm{edge}(\bm{r})$ with the data from our micromagnetic simulations (see Methods). This approach allows us to determine the trajectory immediately and precisely calculate the five forces individually, which helps to understand why the skyrmions reverse their direction of motion under constant current. 

We note that the forces in the Thiele equation are not to be understood in the Newtonian sense since the skyrmion does not move in the direction of the forces' sum, which is zero per definition (cf. equation~\ref{eq:Thiele}). Instead, we have to find the velocity vector, such that all five forces compensate: The reader is reminded that the spin-orbit torque and edge related forces always point along $\pm y$. The skyrmion moves always perpendicular to the gyroscopic force and anti-parallel to the dissipation force. The skyrmion-skyrmion attraction always points towards the center of the sample $\bm{r}=0$ since the two skyrmions move symmetrically.

In the first part of the trajectory [I in figure~\ref{fig:fig3_motionSOT}(b)], the force related to the spin-orbit torque is larger than the force from the edge [cf. figure~\ref{fig:fig3_motionSOT}(c)], as explained before. This means the velocity must be oriented such that an additional gyroscopic force occurs that compensates the spin-orbit force. Since the velocity is always perpendicular to the gyroscopic force, it must have a positive $x$ component. At the reversal point (II), the velocity has drastically decreased since it would be zero if there was no skyrmion-skyrmion interaction due to the compensation of the spin-orbit related force and the edge force, as explained before. However, due to the consideration of the skyrmion-skyrmion interaction, $y_c$ is not a stationary position anymore. The skyrmion consequently moves along $-x$ in (III). This is possible because the dissipative force fully compensates for the interaction force's $x$ component. Additionally, the gyroscopic force must not deliver any $x$ component which is only fulfilled for this direction of motion: If $\bm{v}$ is along $-x$, the gyroscopic force is along $y$. Compared to the discussion without interaction potential, the skyrmion will even move a bit further along $y$, thereby increasing the edge force. The gyroscopic force is oriented along $y$ to compensate for this additional edge force. For such a high value of the $y$ component, all five forces can only compensate each other if the velocity is oriented along $-x$. Upon returning towards $x=0$ the skyrmion-skyrmion interaction becomes stronger. Consequently, the dissipative and gyroscopic forces must also increase so that the skyrmion speeds up.

We close this section by noting that the Thiele equation assumes a rigid skyrmion structure. However, from the micromagnetic simulation, we note that the skyrmion can be deformed, and once it approaches the edge, its size changes. Mathematically this translates into non-diagonal elements of the tensors $\underline{D}$ and $\underline{I}$ different from zero. Since these terms are very small, the idealized trajectory based on the Thiele equation does not differ qualitatively from the trajectory based on the micromagnetic simulation [figure~\ref{fig:fig3_motionSOT}(a)]. However, these non-diagonal components lead to slower resetting dynamics once the skyrmion has reached the edge.

\begin{figure*}[tb]
    \centering
    \includegraphics[width=\columnwidth]{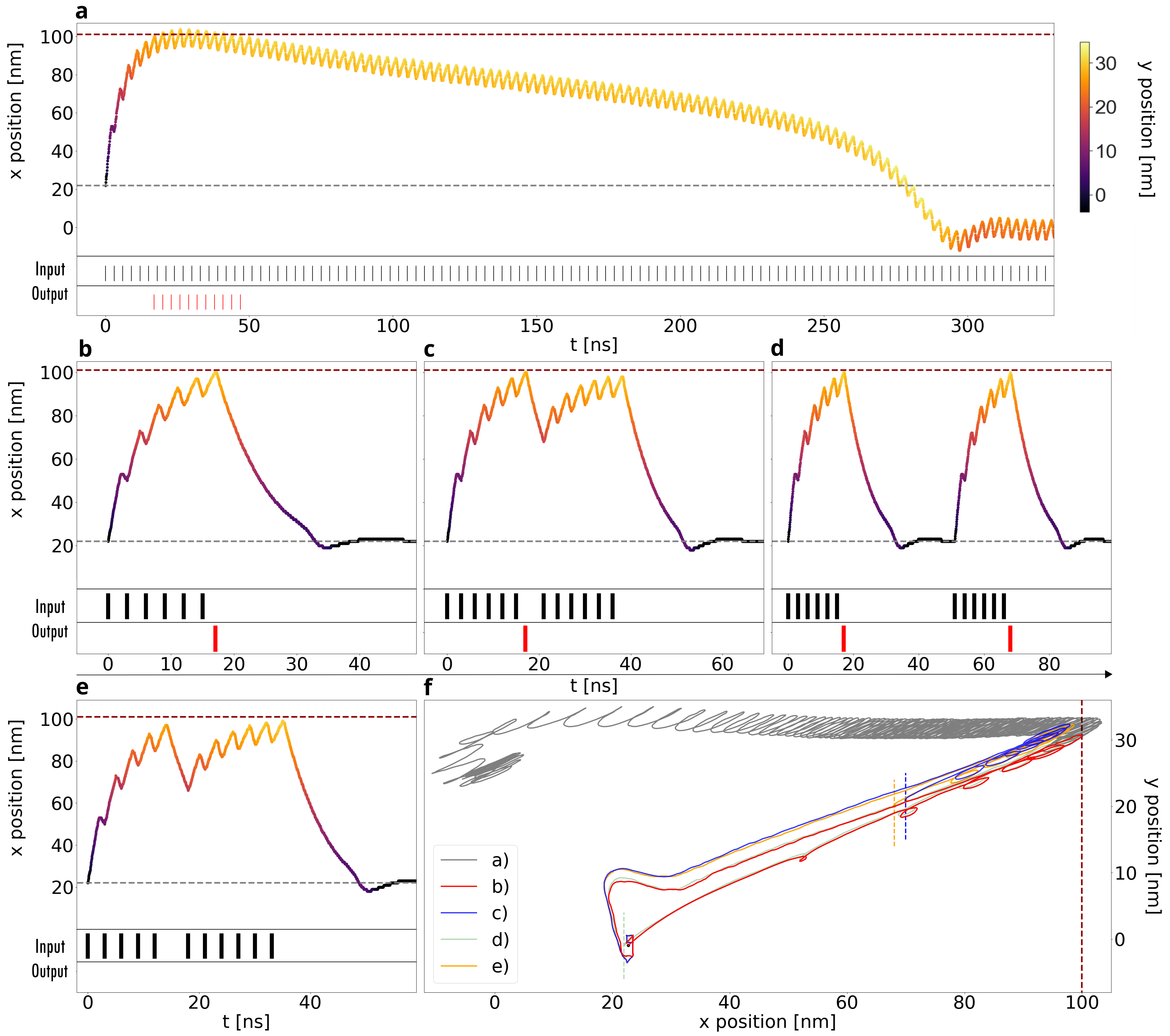}
    \caption{Biskyrmion-based artificial neuron. (a) Equidistant current pulses. The right skyrmion moves similarly to the scenario under constant current presented in figure~\ref{fig:fig3_motionSOT}. The only difference is the occurrence of oscillations with a period of $3\,\mathrm{ns}$ corresponding to the time between two pulses. (b) LIF functionalities. 6 input pulses (black lines) drive the skyrmion towards the detector at $x_d=100\,\mathrm{nm}$ (dashed line) and, due to the skyrmion Hall effect, towards the horizontal edge (color represents the $y$ component). A single output signal (red) is triggered. An animated version is available in the SI. (c) Refractoriness. After the firing event, after 6 pulses, like in a, one pulse is omitted and then another 6 pulses are applied. This time, the neuron does not fire due to an inequivalent $y$ coordinate in comparison to a. (d) End of the refractory period. For this simulation, the time between the two pulse sequences has been increased so that the biskyrmion has reformed in the meantime. The neuron fires once after each pulse sequence. (e) Refractory period without a firing event. This simulation is similar to c, but the first sequence contains only 5 pulses. The neuron does not fire even when a sequence of 6 pulses is applied shortly after. (f) Inequivalence of the trajectories. The trajectories for a, b (equivalent to d), c, and e are compared. The short dashed lines mark the arrival of the second spike train for c, d and e. The long dashed line corresponds to the location of the detector.}
    \label{fig:fig4_pulses_LIF}
\end{figure*}

\subsection{Biskyrmion-based neuronal dynamics.} 
The behavior under current pulses is very similar to the situation explained above. If a continuous sequence of short current pulses is applied, the trajectory looks almost identical to the case with constant currents [figure~\ref{fig:fig4_pulses_LIF}(a)]: The right skyrmion is pushed along the positive $x$ and $y$ directions according to the skyrmion Hall angle. Once it reaches its maximum $x$ coordinate, it slowly moves back along the $-x$ direction towards the center of the sample. However, for a neuron device it is more important to understand the behavior under limited sequences of current pulses, as will be discussed next.

In the following, we will present the neuronal functionalities of the artificial neuron by discussing three archetypal examples. The first case [figure~\ref{fig:fig4_pulses_LIF}(b)] allows to discuss the LIF functionality of our device. We apply a spike train of positive current pulses corresponding to $j\Theta_\mathrm{SH}= 15\,\mathrm{MA}/\mathrm{cm}^2$ and track the skyrmion's core position. The black lines represent the input pulses, each with a duration of $2\,\mathrm{ns}$ applied every $3\,\mathrm{ns}$. Six pulses are sufficient to drive the skyrmion into the detector beginning at $x_d=100\,\mathrm{nm}$ (red dashed line). The red line indicates the output signal fired by the device. 
For this first example, once the artificial neuron has fired, the input signals are turned off, and the skyrmion-skyrmion interaction resets the device. However, compared to the constant current case, the reset period is much shorter since no force is still pushing the two skyrmions apart.

In the second example [figure~\ref{fig:fig4_pulses_LIF}(c)], we present that the biskyrmion adds a layer of bio-fidelity for skyrmion-based artificial neurons by incorporating the refractory signature. To test for refractoriness, $4\,\mathrm{ns}$ after firing, we apply the same input-spike train again. In a device with only the LIF functionality, as in the skyrmion-based neuron from figure~\ref{fig:fig1_neuron_devices}(e), this should be more than enough to trigger another firing event. For the biskyrmion case, however, as shown in figure~\ref{fig:fig4_pulses_LIF}(c), the same spike train does not trigger another firing event. To understand this, we refer to the dynamics under constant current. While the device is being reset, the biskyrmion has not yet formed, and the individual skyrmion has a non-zero position component along the $y$-axis [the colors in figure~\ref{fig:fig4_pulses_LIF}(c)]. The skyrmion moves back on a slightly different path compared to the initial motion towards the edge [figure~\ref{fig:fig4_pulses_LIF}(f)]. Once the second train of pulses arrives, the skyrmion moves again towards the detector but cannot reach the detector because the motion stops at a value $x<x_d$. This type of motion remains unless this hysteresis is resolved by reestablishing the biskyrmion.

We present this particular case in the third archetypal example [figure~\ref{fig:fig4_pulses_LIF}(d)]. This time, the second train of pulses arrives $25\,\mathrm{ns}$ after the firing event so that the biskyrmion has already reformed. The refractory period is overcome and the neuron is fully reset.

Before we conclude, we want to discuss three details about the refractory period. (i) In our simulations, one assumption was that all input current pulses were of the same magnitude. Only under this assumption do we have an absolute refractory period during which it is impossible to make the neuron fire again [similar to the red region of the biological neuron shown in figure~\ref{fig:fig1_neuron_devices}(d)]. If larger input currents are allowed, the neuron can fire again. It is a relative refractory period in this case [similar to the orange region of the biological neuron shown in figure~\ref{fig:fig1_neuron_devices}(d)].
(ii) For the neuron to enter the refractory period, there must be at least one input pulse missing after a firing event. If current pulses keep on being received as inputs, the device will fire several times in a short sequence before it enters the refractory state [figure~\ref{fig:fig4_pulses_LIF}(a)]. Such a signature is called `phasic bursting' and is also present in biological neurons~\cite{izhikevich2004model}.
(iii) Already a sub-threshold input without a firing event can initiate a refractory period in our device [figure~\ref{fig:fig4_pulses_LIF}(e)], which is also in good agreement with the analogous biological neuron.

\section{Conclusions}
In summary, we have predicted an artificial neuron that resembles a biological neuron incorporating the leak, integrate, fire, and refractory characteristics. The goal of artificial neuron devices is to mimic neuronal dynamics with the same speed and power efficiency as the human brain, hardware-wise. So far, mainly the LIF features have been fabricated by major technology companies, for instance, Intel with the Loihi chip~\cite{davies2021advancing, davies2018loihi} and IBM with the TrueNorth chip~\cite{davies2021advancing, davies2018loihi} (for a review of electronic and spintronic artificial neurons, see reference~\cite{zhu2020comprehensive}). We expect that an artificial neuron with a refractory period will help to overcome current performance bottlenecks. 

Our discovery of the unique trajectory of the subskyrmions is also interesting from a fundamental point of view and will bring biskyrmions further into the spotlight of the magnetism community: While a current remains applied along the same direction, the skyrmions revert their direction of motion, caused by the broken rotational symmetry of the biskyrmion. This is the opposite of a skyrmion ratchet~\cite{gobel2021skyrmion}, where the geometry breaks the inversion symmetry to translate an alternating current into a net motion.

\appendix
\section{LIF model}
The dynamics of a neuron without refractory period are analogous to an RC circuit. It can be modeled by a LIF model which has three features: accumulation of the potential (integrate), drop in the potential due to charge leakage (leak), and reaching the threshold value (fire). Mathematically, this is described by 
\begin{equation}
\label{eq:LIF}
    \tau_m \frac{\mathrm{d} u(t)}{\mathrm{d} t} = -( u(t) - U_{0}) + R I(t),
\end{equation}
where $\tau_m = RC$ is a time constant defined by the membrane resistance $R$ and the capacitance $C$. 

In figure~\ref{fig:fig1_neuron_devices}(e), we plot the solution of equation~\ref{eq:LIF} under a periodic input of $2$ units duration every $3$ units of time. The membrane potential reaches the threshold $U_m$ after six inputs with the appropriate choice of parameters with $\tau_m = 0.18, U_0 = 10$ and $R =1$ (dimensionless units). After firing, the input is turned off, and the membrane potential drops towards $U_0$.

\section{Potentials for the Thiele equation.} 
\begin{figure*}[tb]
    \centering
    \includegraphics[width=\columnwidth]{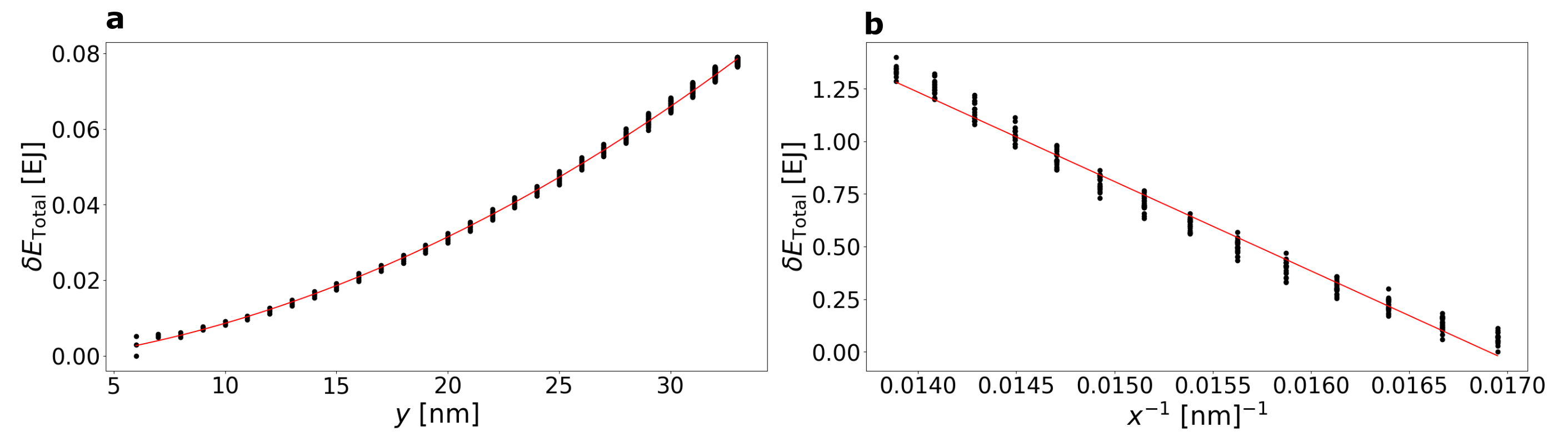}
    \caption{Total energy of the two simulated scenarios explained in the Appendix. In (a), the only non-constant contribution to the total energy is $U_\mathrm{edge}$. In (b), the only non-constant contribution to the total energy is $U_\mathrm{int}$. The difference of the $x$ coordinates has been plotted as the reciprocal value to be able to fit a linear function.}
    \label{fig:fit}
\end{figure*}

In order to simulate the motion of the two subskyrmions based on the Thiele equation (equation~\ref{eq:Thiele}), we have determined the tensors and potentials via fitting data from our micromagnetic simulations. We find that the skyrmion-edge interaction follows $U_\mathrm{edge}(y) \approx \lambda_2 y^2 $ and that the skyrmion-skyrmion interaction follows $U_\mathrm{int}(\bm{r}_{12}) \approx k_1/r_{12} $ where $\lambda_2$ and $k_1$ are the strength of the edge and skyrmion-skyrmion potentials, respectively. For the numerical calculation, these coefficients are obtained from two different micromagnetics simulations. For $U_\mathrm{edge}$, we simulate only one Bloch skyrmion being pushed towards the edge by SOT and fitted the energy versus $y$ position [figure~\ref{fig:fit}(a)]. For $U_\mathrm{int}(\bm{r}_{12})$, we write two Bloch skyrmions $100$ nm apart from each other along the $x$ direction to avert the edge interaction. Here, no current is induced and only the skyrmion-skyrmion interaction moves the skyrmions towards each other. The shape of the potentials and the coefficients are obtained by fitting energy versus skyrmion position [figure~\ref{fig:fit}(b)].
The resulting fit is
\begin{equation}
    U_\mathrm{edge}(y)= \lambda_2 y^2 + \lambda_0,\quad\quad\quad\quad
    U_\mathrm{int}(x_{12})= \frac{k_1}{x_{12}} + k_0,
\end{equation}
where $\lambda_2 \approx 5.9\times10^{-13}$ EJ/m$^2$, $\lambda_0 \approx -16.18$ EJ, $k_1 \approx -0.425\times 10^{-9}$ EJ/m and $k_0 \approx -15.77$ EJ.

For the tensors, we have fitted a skyrmion right after the motion has started: $D_{xx}=D_{yy} \approx 14$, $I_{xx}=I_{yy}\approx98$ nm.

\section*{Data availability}
Data that support the findings of this work are available from the authors on reasonable request.

\section*{Code availability}
For the micromagnetic simulations we used the open-source code mumax3 available at https://mumax.github.io/.

\section*{Acknowledgements}
This project has received funding from the European Union’s Horizon 2020 research and innovation programme under the Marie Skłodowska-Curie grant agreement No 955671. This work is supported by SFB TRR 227 of Deutsche Forschungsgemeinschaft (DFG).

\section*{Author contributions}
I.A. performed the simulations with the help of B.G.
B.G. and I.A. wrote the manuscript with significant inputs from all authors.
I.A. prepared the figures.
All authors discussed the results.
B.G. and I.M. planned and supervised the project.

\section*{Supplementary information}
Animated figures accompany this paper at [insert link].

\section*{Competing interests}
The authors declare no competing interests.

\printbibliography
\end{document}